\documentclass[aps,pre,twocolumn,showpacs,superscriptaddress,groupedaddress]{revtex4-1} 

\usepackage{graphicx}        
\usepackage{dcolumn}        
\usepackage{bm}             
\usepackage{amssymb}
\usepackage{amsmath}  
\usepackage{color}
\definecolor{korr_26Apr}{rgb}{0,0,0} 
\definecolor{red}{rgb}{1,0,0}

\begin{document}

\widetext

\title{Electric Field and Humidity Trigger Contact Electrification}
\author{Yanzhen Zhang$^1$}
\author{Thomas P\"ahtz$^{2,3}$}
\email{0012136@zju.edu.cn}
\author{Yonghong Liu$^1$}
\email{Liuyhcup@163.com}
\author{Xiaolong Wang$^1$, Rui Zhang$^1$, Yang Shen$^1$, Renjie Ji$^1$ \& Baoping Cai$^1$}
\affiliation{1.~College of Electromechanical Engineering, China University of Petroleum, 266580 Qingdao, China \\
2.~Institute of Physical Oceanography, Ocean College, Zhejiang University, 310058 Hangzhou, China \\
3.~State Key Laboratory of Satellite Ocean Environment Dynamics, Second Institute of Oceanography, 310012 Hangzhou, China}

\begin{abstract}
Here, we study the old problem of why identical insulators can charge one another on contact. We perform several experiments showing that, if driven by a preexisting electric field, charge is transferred between contacting insulators. This happens because the insulator surfaces adsorb small amounts of water from a humid atmosphere. We believe the electric field then separates positively from negatively charged ions prevailing within the water, which we believe to be hydronium and hydroxide ions, such that at the point of contact, positive ions of one insulator neutralize negative ions of the other one, charging both of them. This mechanism can explain for the first time the observation made four decades ago that wind-blown sand discharges in sparks if and only if a thunderstorm is nearby.
\end{abstract}

\maketitle

\section{Introduction}
Contact electrification, which describes the phenomenon that two contacting insulators can acquire electric charges when they are separated, is responsible for numerous mysterious natural phenomena, such as the generation of electrified particles in wind-blown sand and dust \cite{Shinbrot_and_Herrmann_2008,Lacks_2010,Kok_and_Renno_2008,Merrison_2012}, lightning near volcanic dust plumes \cite{McNutt_and_Davis_2000}, and devastating explosions in grain and coal plants \cite{Abbasi_and_Abbasi_2007}. Contact electrification is also extensively exploited in industrial applications, such as electrophotography \cite{Schein_2009}, Laser printing \cite{Polsen_et_al_2013}, 3D printing \cite{Kawamoto_2007,Calvert_2001} and electrostatic separations \cite{Kwetkus_1998}. Despite this huge importance of contact electrification for nature and industry, its underlying physics remain elusive \cite{Castle_2008,McCarty_and_Whitesides_2008,Lacks_and_Sankaran_2011,Pahtz_et_al_2010,Schein_2007}, even though it has been studied since ancient Greece. The main difficulty is to understand how insulators, which by definition have no free charge carriers, can charge one another on contact. In fact, many complex contact electrification mechanisms, based on electron transfer \cite{Liu_and_Bard_2008,Diaz_and_Felix-Navarro_2004}, ion transfer \cite{McCarty_and_Whitesides_2008}, transfer of charged material \cite{Baytekin_et_al_2011a}, asymmetric partitioning of hydroxide ions \cite{Gu_et_al_2013,Ducati_et_al_2010}, and nanochemical reactions \cite{Williams_2012a,Burgo_et_al_2013} have been proposed, but the scientific debate remains controversial \cite{Williams_2012b,Knorr_2011,Baytekin_et_al_2011a}. For instance, charge transfers between insulators often depend on the contact mode (e.g., point contact, area contact, rubbing contact) and other specified conditions \cite{Williams_2012b}, with the consequence that certain mechanisms may be predominant for certain conditions, but negligible for other conditions.


Moreover, certain phenomena, which are associated with contact electrification, remain insufficiently explained: For instance, in 1971 Kamra made the fascinating observation of electric activity when a thunderstorm blew over gypsum sand dunes in New Mexico \cite{Kamra_1972}. On the top of several dunes, he saw sparks without branches extending from the ground straight up to a few meters height in the air and clearly distinguished them from thunderstorm lightning. However, at days without nearby thunderstorm, but similarly strong winds, Kamra \cite{Kamra_1972} mysteriously did not observe any such sparks. Recently, Ref. \cite{Pahtz_et_al_2010} also made similar observations in the laboratory: The authors fluidized a particle bed filling a glass jar. With fluidization the bed particles started to collide with each other. They then acquired large electric charges when a pre-existing electric field generated by a van de Graaff generator was applied on the jar, but did not so when the generator was turned off. The relative air humidity in these experiments and Kamra's observation was of comparable magnitude. Ref. \cite{Pahtz_et_al_2010} proposed and simulated a charge transfer mechanism to explain their laboratory experiments: Under the presence of a sufficiently strong pre-existing electric field, oppositely charged charge carriers gather at opposite hemispheres of the insulators before collision, so that at a collision of two insulators the charge carriers within the colliding hemispheres neutralize each other, which charges both colliding particles. The simulations further considered that particles are neutralized when they hit the grounded bottom, which allowed net-charging the agitated particle cloud, even though the charge transfer mechanism conserves charge. These simulations were in quantitative agreement with the laboratory observations, and thus offered a possible explanation for Kamra's \cite{Kamra_1972} observations for the first time \cite{Lacks_2010}. However, since Ref. \cite{Pahtz_et_al_2010} neither specified the identity of the involved charge carriers nor the manner in which they move within the insulator, a complete picture explaining Kamra's \cite{Kamra_1972} observations has still been missing.

Here we investigate a contact electrification mechanism, which is triggered by the interplay of a pre-existing electric field ($E$) and humidity, by experimental means. As we explain in the following sections, we believe this mechanism works in the following way: First, the insulators adsorb water from a humid environment on their surfaces. Then positively and negatively charged ions dissolved in this water, we believe hydronium and hydroxide ions, polarize and thus gather at opposite sides of the insulators due to the electric field. On contact they neutralize the respective other species around the contact domain. Finally, when separated, one insulator remains with most of the positively and the other one with most of the negatively charged ions. Apparently, the here described charge transfer along and between the insulators resembles the aforementioned charge transfer mechanism of Ref. \cite{Pahtz_et_al_2010}. However, since also the identity of the involved charge carriers and the manner in which they move along the insulator surface are specified, our studies offers a first complete explanation for the aforementioned observations of Kamra \cite{Kamra_1972}.

Our study indicates that, at least for point contacts (e.g., interparticle collisions), the mechanism we investigated seems to be much stronger than other contact electrification mechanisms if the electric field and the relative humidity are sufficiently large. Interestingly, humidity and thus hydronium and hydroxide ions due to the dissociation of water have been associated with a large variety of electrostatic effects and even been identified as key components in contact electrification in many previous studies \cite{McCarty_and_Whitesides_2008,Burgo_et_al_2011,Ducati_et_al_2010,Wiles_et_al_2004,Gouveia_and_Galembeck_2009,Ducati_et_al_2010,Pence_et _al_1994, Baytekin_et_al_2011b,Gu_et_al_2013,Zheng_et_al_2014}. Humidity has also been linked to the pick-up of insulating particles from a grounded particle bed \cite{Kok_and_Renno_2006} or from a grounded conductive plane \cite{Sow_et_al_2013} by pre-existing electric fields. In fact, in laboratory experiments such particles were lifted by strong electric fields in the presence of sufficient humidity. It was speculated that this happens because the particles charge via induction (i.e., the redistribution of electrical charge in a conductor due to electric fields) due to water on their surfaces adsorbed from a humid environment. Indeed, the water makes these surfaces conductive, resulting in an upwardly-directed lifting force due to the electric field \cite{Lebedev_Skalskays_1962}. Note that this speculated lift mechanism strongly resembles our contact electrification mechanism. The important difference is the contact time: It is not a priori clear that the lifting mechanism, in which particles are in enduring contacts with their conductive surroundings before they are lifted from the surface, can be generalized to a contact electrification mechanism working for general contacts of insulators, including comparably very brief contacts during interparticle collisions in midair. Moreover, it is actually unclear why adsorbed water on the particle surfaces makes them conductive because experimental studies have shown that water films on particle surfaces may not be continuous, but instead covered by a multitude of small water islands, even for very hydrophilic particles \cite{Xu_et_al_2010,Carrasco_et_al_2012}. In our study, we propose that adsorbed water may make the insulator surfaces quasi-conductive even if the water films are discontinuous because charge might be exchanged between these water islands and water vapor surrounding the insulators.

\section{Experiments within silicone oil}
\begin{figure}
 \begin{center}
  \includegraphics[width=1.0\columnwidth]{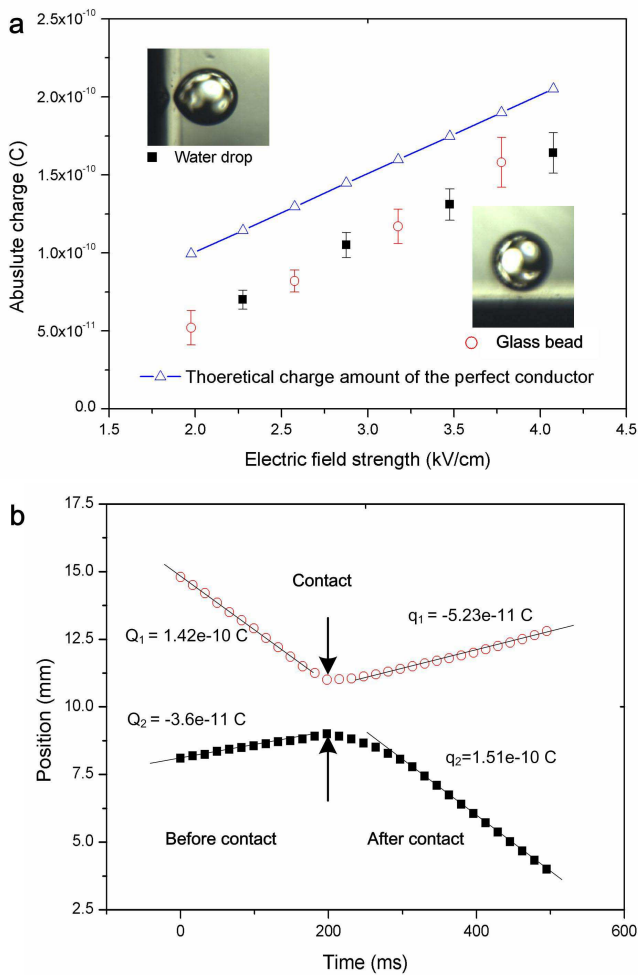}
 \end{center}
 \caption{Experiments within silicone oil: (a) Absolute value of the charges of a glass bead ($r=1$mm) and a water drop ($r=1$mm), bouncing between two charged metal electrodes, after contact with the electrodes versus electric field strength (symbols). The charges were estimated from a balance between the viscous drag and Coulomb forces, while the solid line displays the theoretical charge of a perfectly conducting sphere ($r=1$mm) \cite{Ristenpart_et_al_2009}. (b) Positions relative to the cathode and charges of two glass beads colliding between two charged metal electrodes ($E=2.5$kV$/$cm) before and after their collision ($Q_1+Q_2\approx q_1+q_2$).}
 \label{Singledoublebead}
\end{figure}

\begin{figure}
 \begin{center}
  \includegraphics[width=1.0\columnwidth]{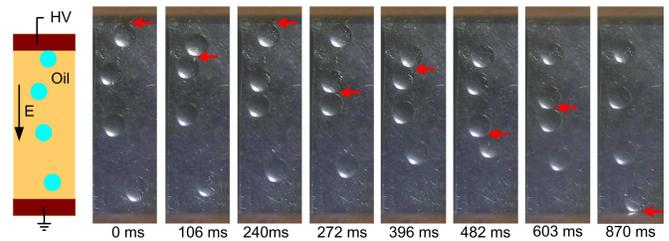}
 \end{center}
 \caption{Experiments within silicone oil: Four glass beads ($r=1$mm) bouncing between two charged metal electrodes ($E=2.5$kV$/$cm). The arrows indicate the locations of charge exchange during each bounce. All pictures show the view from the top of the cell.}
 \label{Fourbead}
\end{figure}

\begin{figure}
 \begin{center}
  \includegraphics[width=1.0\columnwidth]{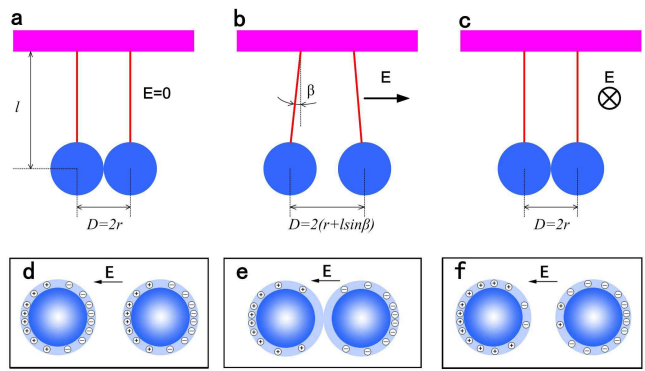}
 \end{center}
 \caption{Experiments within silicone oil (a-c): Two contacting glass beads ($r=1$mm) placed on fibers hanging down a bar. (a) Separation of the glass beads did not occur in the absence of an electric field. (b) Separation did occur under the influence of an electric field ($E=2.5$kV$/$cm) directed normal to the contact plane. (c) Separation did not occur under the influence of an electric field directed tangential to the contact plane. (d-f) Charge transfer mechanism of hydrophilic particles in contact with each other. Note that the water film was amplified and drawn as continuous for illustration. The real water film is very thin and may not be continuous.}
 \label{Efield}
\end{figure}

\begin{figure}
 \begin{center}
  \includegraphics[width=1.0\columnwidth]{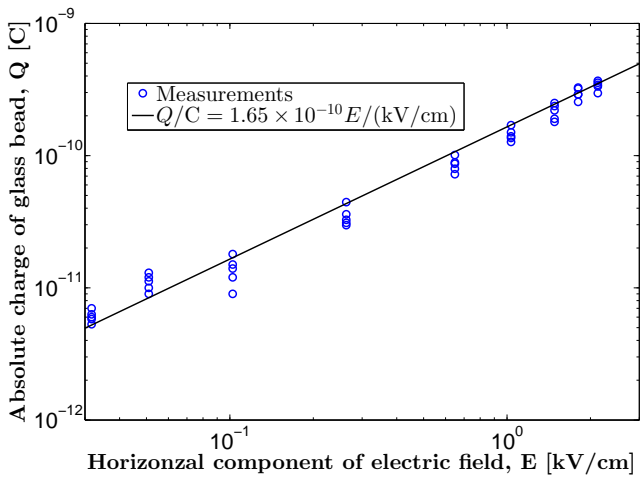}
 \end{center}
 \caption{Experiments within silicone oil: Two contacting glass beads ($r=1$mm) placed on fibers hanging down a bar. Measurements of the absolute value of the charge of the glass beads as a function of the electric field component in direction normal to the contact plane after removal of one of the beads.}
 \label{EQ}
\end{figure}

We carried out experiments within cells completely filled with silicone oil. These experiments are described below (a detailed description of these experiments can be found in Appendices A and B):
\begin{enumerate}
 \item We placed a glass bead between two charged metal electrodes and observed that it bounced forth and back between these electrodes. We measured the absolute value of the charge of the glass bead after contact with these electrodes as a function of $E$ (see Fig.~\ref{Singledoublebead}a). These experiments confirm that charging of the glass bead occurs during the contact with these electrodes.
 \item We placed either two or four identical glass beads between these electrodes and observed that they bounce between each other and the electrodes (see Figs.~\ref{Singledoublebead}b and \ref{Fourbead}, and Supplementary Movie 1). In the case of two glass beads, we measured the charges of both beads before and after a collision (see Fig.~\ref{Singledoublebead}b). These experiments confirm that a large amount of charge is transferred between the two glass beads at contact since the total charge of the two beads remains roughly constant.
 \item While in experiments~1 and 2 the glass beads were stored in air before placed into the silicone oil, we now baked them for several hours to get rid of most of the water on their surfaces. After baking no bouncing was observed, confirming that the presence of water on the surface of the glass beads is necessary for the their charging.
 \item We hung two fibers down a bar. At the end of each fiber a glass bead was placed, and the distance between the two fibers was exactly one bead diameter, meaning that the glass beads were initially just in contact. If we activated a strong electric field ($E=2.5$kV/cm) directed normal to the contact plain, the glass beads separated (cf. Fig.~\ref{Efield}b), while no separation occurred if the electric field was absent (cf. Fig.~\ref{Efield}a) or directed parallel to the contact plane (cf. Fig.~\ref{Efield}c). These experiments confirm that a pre-existing electric field is responsible for the charging of the glass beads, and that the relevant component is the one in direction normal to the contact plane. Moreover, we now varied the component of $E$ in direction normal to the contact plane and then, regardless of whether the glass beads separated or not, manually removed one bead including its fiber from the setup after turning on the field. After the removal, we increased the electric field to a high value ($E=3.75$kV/cm) and measured the absolute value of the charge of the remaining bead indirectly from its displacement due to the Coulomb force. We found that it is approximately proportional to the electric field component in direction normal to the contact plane (see Fig.~\ref{EQ}).
\end{enumerate}


\section{Identity of charge carriers}
What are the charge carriers giving the glass beads their charges? The fact that the glass beads must have been subjected to a humid atmosphere prior to the experiments to receive charges in contacts indicates that the surfaces of the glass beads have adsorbed water from the atmosphere (see Appendix D for different adsorption mechanisms). Hence, the charge carriers are either ions dissolved in the adsorbed water or charge carriers produced by reactions between the insulator surface and substances dissolved within the adsorbed water (for instance, O$_2$ dissolved in the adsorbed water can considerably increase the surface conductivity of undoped diamond \cite{Chakrapani_et_al_2007}). To minimize the likelihood of chemical reactions with the glass bead surface, we redid the experiments~1 and 2, but this time we subjected the glass beads to a humid N$_2$ atmosphere before placing them into the silicone oil. In fact, we observed exactly the same bouncing behaviors as before. Since N$_2$ is a very inert gas and thus very unlikely to react with the surface of the glass beads, ions dissolved in the adsorbed water remain as the most likely charge carrier candidates. This is further supported by our measurements showing that the absolute charge of the insulator after contact with the electrodes as a function of $E$ resembles that of a water drop (see Fig.~\ref{Singledoublebead}a), indicating that the contact electrification mechanism is not related to chemical properties of the glass bead. However, one should be aware that this resemblance might be coincidental. Concerning the identity of the ions dissolved in the adsorbed water, we believe, like many previous studies under similar circumstances \cite{McCarty_and_Whitesides_2008,Burgo_et_al_2011,Ducati_et_al_2010,Wiles_et_al_2004,Gouveia_and_Galembeck_2009,Ducati_et_al_2010,Pence_et _al_1994}, that the majority of them are hydronium and hydroxide ions due to the reaction 2H$_2$O$\rightleftharpoons$H$_3$O$^+$+OH$^-$, especially in the cases in which the glass beads where stored in a humid N$_2$ atmosphere. This is because in water saturated with nothing but N$_2$, other ions seem unlikely to be created by standard chemical reactions. Nonetheless, it cannot be excluded that other ions, such as HCO$_3^-$ due to remnants of CO$_2$ dissolved in the water or ions from substrates with ion bonds contaminating the insulator surface, are significantly or even predominantly involved.

\section{Charge transfer mechanism}
Knowing that the charge carriers are most likely ions dissolved in water adsorbed on the insulator surface, it remains to answer the question what happens to these ions in an electric field before, at and after contact. Assuming that these charge carriers can be transported along the insulator surface, the electric field, if sufficiently strong, will make the negatively (positively) charged ions gather at the hemisphere closest (farthest) from the anode (cf. Fig.~\ref{Efield}d). At contact, presuming the electric field is normal to the contact plane, the positively charged ions of one glass bead neutralize the negatively charged ions of the other one within a small water bridge forming around the contact point (cf. Fig.~\ref{Efield}e). When the glass beads separate, one glass bead remains with most of the positively and the other one with most of the negatively charged ions (cf. Fig.~\ref{Efield}f).

The charge transfer mechanism described above requires that the ions can be transported along the insulator surface. One possible transport mechanism is that the ions can freely move in a continuous water film coating the insulator surface. However, this possibility seems unlikely since experiments have shown that water films on particle surfaces may not be continuous, even for very hydrophilic particles \cite{Xu_et_al_2010,Carrasco_et_al_2012}. Instead particle surfaces might be covered by a multitude of small water islands \cite{Xu_et_al_2010,Carrasco_et_al_2012}. We thus speculate that the ions instead might hop from water island to water island. This might be possible because each water island might be able to exchange charges with the surrounding water vapor. In fact, charge transfer between metal surfaces and surrounding water vapor under high humidity have been confirmed experimentally \cite{Ducati_et_al_2010}. Since the lifetime of adsorbed water molecules is of the order of milliseconds \cite{Ewing_2006}, there should always be water vapor surrounding the particles, even in our silicone oil experiments. 

\section{Sensitivity of charge transfer to surface material properties}
\begin{figure}
 \begin{center}
  \includegraphics[width=1.0\columnwidth]{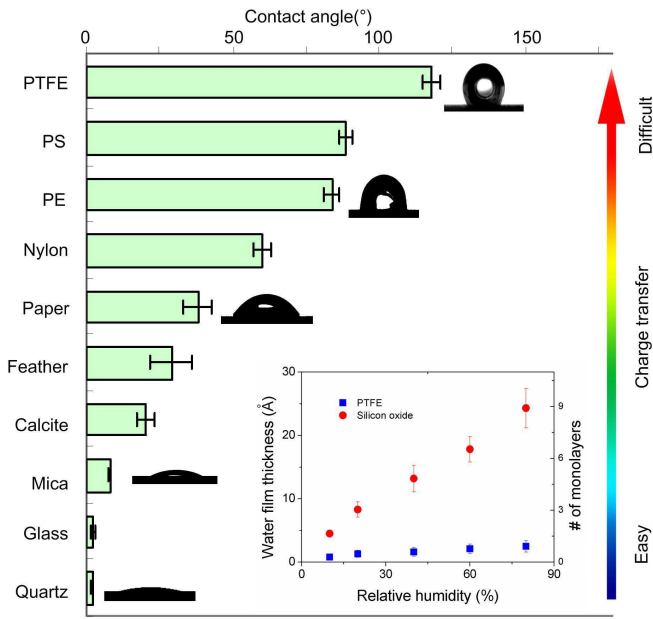}
 \end{center}
 \caption{Difficulty of charge transfer (qualitatively) versus contact angle between a 1$\mu$L water drop and the insulator surface for several dielectric materials measured with an Optical Contact Angle Measuring Device. Inset: Number of monolayers of water molecules composing the water films on the surfaces of silicon oxide and PTFE, measured with an ellipsometer, versus air humidity.}
 \label{Sensitivity}
\end{figure}

Whatever the exact charge carrier transport mechanism might be, it is at least clear that it is the more efficient the more water is available on the insulator surface. This suggests that the aforementioned charge transfer mechanism works better for hydrophilic materials, such as glass, than for hydrophobic materials, such as polyethylene (PE), polystyrene (PS) and Polytetrafluoroethene (PTFE). Hydrophilic materials are characterized by large free surface energies, allowing them to form attractive bonds with the water molecules \cite{Ewing_2006}. We estimated the free surface energy of the insulators indirectly by measuring the contact angle between a $1\mu$L water drop and the insulator surface with large contact angles corresponding to small free surface energies. Indeed, we found that the larger was the contact angle the more difficult was the charge transfer between contacting insulators or between insulators and electrodes (see Fig.~\ref{Sensitivity}). We concluded this from quantitative measurements of the absolute charge of spherical beads after contact with the electrodes (see Appendix C) and from qualitative observations of the bouncing behavior of objects (see Supplementary Movie 2) for different dielectric materials: For instance, bouncing between the electrodes did not occur for the most hydrophobic materials PE, PS and PTFE. Also, when PE and PS particles came in contact with the electrodes, they remained in contact and rolled on the electrodes for a while before eventually separating (see Appendix E for a possible mechanism). PTFE particles, which are more hydrophobic than PE and PS, did not separate at all from the electrodes after contact. In fact, by measuring the average thicknesses of the water covering the surfaces of PTFE and silicon oxide particles (as a representative for hydrophilic materials) as a function of air humidity, we found that the average water thickness on the surface of PTFE particles is between one and two monolayers of water molecules almost independent of the air humidity (inset of Fig.~\ref{Sensitivity}). This is very probably too thin to allow efficient ion transport and explains why contact electrification for PTFE particles was much weaker than for hydrophilic particles. The ability of a PTFE particle to adsorb water can be increased by immersing it in saturated sodium dodecyl benzene sulphonate (SDBS) ethanol solution. After doing so, the contact angle between the 1$\mu$L water drop and the particle surface decreased from about $120^\circ$ to about $15^\circ$, indicating that a hydrophilic molecular layer with a large free surface energy assembled on the PTFE particle surface. Our measurements show that, after surface modification, the PTFE particles bounced between the electrodes, just like hydrophilic particles (see Supplementary Movie 3).

\section{Conclusion}
What is the relevance of the contact electrification mechanism we investigated in comparison to previously described mechanisms? To answer this, we emphasize that the absence of bouncing of glass beads between the electrodes in experiments~1-3 or undetectable glass bead displacement in experiment~4. do not necessarily imply the absence of contact electrification, for instance due to previously described mechanisms, because the charging may have been too weak to be noticed in our experimental setups. However, they do imply that the charging is much weaker in comparison to the cases in which bouncing occurs and glass bead displacement is significant, respectively. This means that, at least for point contacts (e.g., interparticle collisions), the mechanism we investigated seems to be much stronger than other mechanisms if the electric field and the relative humidity are sufficiently large. This particular conclusion is supported by a recent study on contact electrification \cite{Siu_et_al_2014}, which experimentally investigated the build-up of electric fields in an agitated particle bed in the absence of a pre-existing electric field. This study reported that the agitated bed does not build up an electric field if the relative humidity is too large ($>45\%$). Assuming the correctness of our interpretation of how our contact mechanism works, this can be easily explained since our contact electrification mechanism tends to dissipate existing electric fields. This means that any attempt of the agitated bed to build up an electric field is countered by the tendency of the field to be dissipated due to our mechanism. This tendency is the stronger the larger the humidity, explaining the existence of a critical humidity beyond which electric fields cannot be built up anymore.

Moreover, it has been shown that the kind of charge transfer mechanism we describe in Figs.~\ref{Efield}d-f can lead to a huge build-up of charge in wind-blown particle clouds if the pre-existing electric field is sufficiently strong \cite{Pahtz_et_al_2010}. In fact, even though this charge transfer mechanism conserves the total charge, a charge build-up can occur due to settling down and neutralization of charged particles at the surface \cite{Pahtz_et_al_2010}. This might explain why Kamra observed sparks at the top of gypsum dunes if and only if a thunderstorm was nearby \cite{Kamra_1972,Lacks_2010}: In fact, the thunderstorm provided a sufficiently strong electric field and humidity (Kamra reported $E\approx0.04$kV/cm and $36-58\%$ humidity) triggering our contact electrification mechanism. Since, according to our interpretation, this mechanism is based on the charge transfer mechanism described in Figs.~\ref{Efield}d-f, it led to the aforementioned huge charge build-up and thus to highly electrified particles, which eventually discharged in sparks.

Finally, the fact that the charging of hydrophilic insulators in electric fields seems to resemble that of water drops (see Fig.~\ref{Singledoublebead}a) might allow controlling contact electrification by controlling the electric fields and the compositions of the insulator surfaces. Exploiting this fact might open new opportunities for industrial applications of contact electrification in the future.

\section*{Appendices}
\renewcommand{\thefigure}{A\arabic{figure}}
\setcounter{figure}{0}
\subsection{Detailed description of experiments~1, 2, and 3.}
\begin{figure}
 \begin{center}
  \includegraphics[width=1.00\columnwidth]{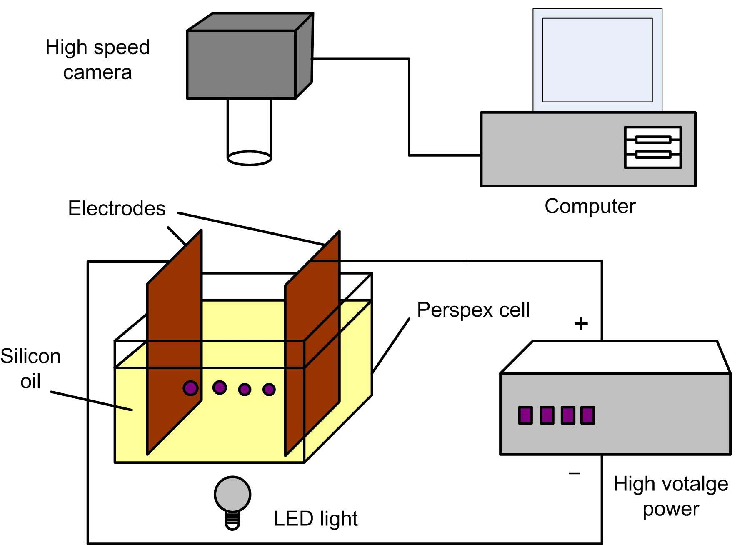}
 \end{center}
 \caption{Illustration of experiments~1, 2, and 3.}
 \label{Sketch1}
\end{figure}

The experimental setup used in experiments~1, 2, and 3 is illustrated in Fig.~\ref{Sketch1}. Two identical copper electrodes were placed in a Perspex cell filled with silicone oil. The width, height, and thickness of the electrodes were $16$mm, $20$mm, and $3$ mm, respectively. The distance between the electrodes was $16$ mm, the viscosity ($\mu$) of the silicone oil was $100$ mPa.s at 25$^\circ$C. The high voltage power can be set to values between $0$ and $10$kV, therefore the electric field ($E$) can be set to values between $0$ and around $6.25$kV/cm. A high speed camera was used to record the bouncing movements of the glass beads or the water drops between the electrodes. A LED cool light was used in order to avoid heating up the silicone oil. In the water drop experiments (see Fig.~\ref{Singledoublebead}a), a pipette was used to generate water drops with a radius ($r$) of $1$mm, which is equal to the radius of our glass beads. All the experimental runs were done at room temperature (25$^\circ$C).

An experimental run was started by turning on the electric field and then placing the glass bead(s) or the water drop(s), respectively, one by one, but in fast sequence without interruption, onto the surface of the silicone oil. They sink in and it takes around $1s-2s$ time until the glass beads hit the bottom, which is when we started capturing their movement with our camera. During the experimental run, the glass beads slide or roll along the bottom and thereby bounce between the electrodes and each other. In experiments~1 and 2, the glass beads were exposed to air with $50\%$ relative humidity for two days before the experimental run was started, while they were baked in experiment~3. The initial horizontal movement of the glass beads in experiments~1 and 2, before they had their first collision with the electrode or another bead, is due to a small amount of excess charge of the adsorbed water \cite{Rezende_et_al_2009}. Due to the absence of this charged water in experiment~3, we put the glass beads initially in contact with the electrodes and/or with each other, but no horizontal movement occured at all, and all particles settled down at the bottom after a few seconds.

From the images captured by the camera, we determined the velocities of the glass beads or the water drops, respectively. Once obtained these velocities were further used to determine the charge of the glass beads or the water drops, respectively, as we explain in the following: Since silicone oil is a highly viscous fluid and the velocities of the glass beads are small (the oil was at rest), we can calculate the fluid drag force ($\mathbf{F_d}$) on a glass bead using Stokes law,
\begin{eqnarray}
 \mathbf{F_d}=-n\pi\mu r\mathbf{U}, \label{Stokes}
\end{eqnarray}
where $\mathbf{U}$ is the velocity of the glass bead and $n$ a shape factor which takes into account that the glass bead is not a perfect sphere ($n=6$ for a perfect solid sphere). We determined $n=6.12$ from a single glass bead falling with constant settling velocity within the silicone oil ($G-F_b+n\pi\mu rU_z=0$, where $G$ is the magnitude of the gravitational force on the glass bead, $F_b$ the magnitude of the buoyancy force on the glass bead, and $z$ the vertical direction). It has been shown that Eq.~(\ref{Stokes}) is also fulfilled for water drops, even though they are liquid, with a coefficient of $n\approx4$ \cite{Im_et_al_2012,Im_et_al_2011}. Indeed, from a single water drop falling with constant settling velocity within the silicone oil, we determined $n=3.91$. Knowing $n$, we determined the charge ($Q$) of the glass bead or the water drop, respectively, within our electric field ($E$) from the electrodes through $QE-n\pi\mu rU_x=0$, where $x$ is the horizontal coordinate, as
\begin{eqnarray}
 Q=\frac{n\pi\mu rU_x}{E},
\end{eqnarray}
whereby we neglected the frictional force from the sliding or rolling motion on the bottom because it is typically much smaller than the drag force. We indirectly confirmed this by performing an experimental test run in which the glass beads bounced vertically and thus without wall friction instead of horizontically. In this test run, we measured virtually the same charges.

\renewcommand{\thefigure}{B\arabic{figure}}
\setcounter{figure}{0}
\subsection{Detailed description of experiment~4.}
\begin{figure}
 \begin{center}
  \includegraphics[width=1.00\columnwidth]{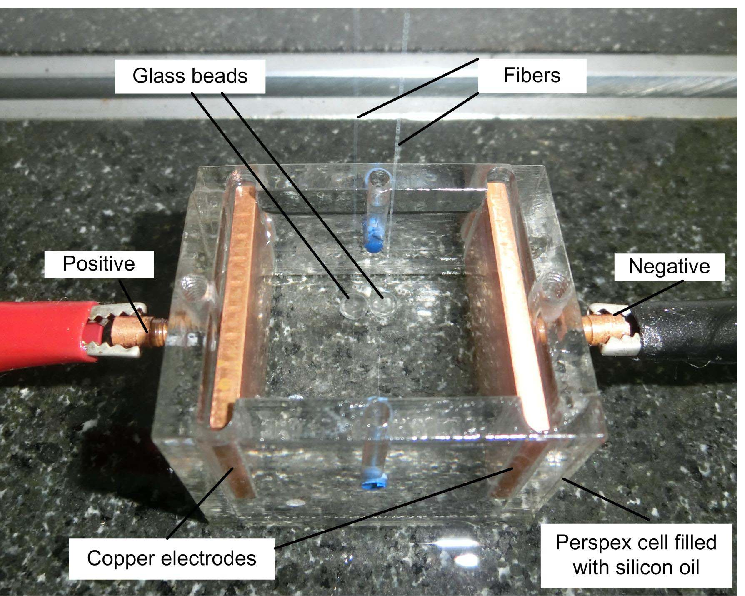}
 \end{center}
 \caption{Illustration of experiment~4.}
 \label{Sketch2}
\end{figure}

The experimental setup used in experiment~4 is illustrated in Fig.~\ref{Sketch2}. Two identical glass beads ($r=2$mm) were hang down a bar using fibers with $10\mu$m diameter and $55$mm length. The glass beads were initially just in contact and put between two identical copper electrodes within a Perspex cell filled with silicone oil. The width, height, and thickness of the electrodes were $20$mm, $25$mm and $3$mm, respectively. The distance between the electrodes was $32$mm, such that the electric field could be set to values between $0$ and about $4$KV/cm. A camera was used to record the position of the glass beads during the experiments. The glass beads were exposed to air with $50\%$ relative humidity for two days before the experimental run was started. All the experimental runs were done at room temperature (25$^\circ$C).

\begin{figure}
 \begin{center}
  \includegraphics[width=1.00\columnwidth]{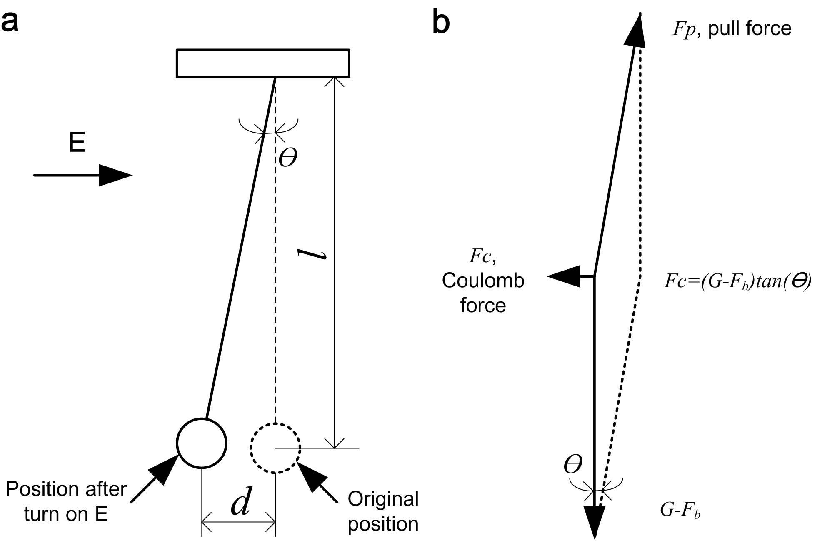}
 \end{center}
 \caption{Sketch of glass bead displacement due to electric field.}
 \label{Sketch3}
\end{figure}

An experimental run was started by turning on the electric field. If the electric field component in direction normal to contact plane ($E)$ was stronger than about $2$kV/cm, the two glass beads separated from each other, otherwise they did not separate (cf. Figs.~\ref{Efield}a-c) because the attractive force (cohesion and/or adhesion) between the two beads was too strong. Still, even if they did not separate, they acquired electric charge whose value we determined as function of $E$ (see Fig.~\ref{EQ}) in the following way: Regardless of whether the glass beads separated or not, we manually removed one bead including its fiber from the setup after turning on the field. Afterwards we increased $E$ to a high value ($3.75$kV/cm) and measured the displacement ($d$) of the bead due to the Coulomb force by comparing the camera pictures with and without electric field. From $d$ we obtained the charge indirectly from the force balance
\begin{eqnarray}
 QE=F_c=(G-F_b)\tan(\theta)
\end{eqnarray}
(see Fig.~\ref{Sketch3}) as
\begin{eqnarray}
 Q=\frac{(G-F_b)\tan\arcsin(d/l)}{E}.
\end{eqnarray}
We note that, if the electric field was weaker than $0.2$kV/cm, we used longer fibers ($215$mm) in order to better detect the displacement of the glass bead. Furthermore, we wish to emphasize that we verified using the same method that the glass beads did not charge if they were not initially in contact with each other.

\renewcommand{\thefigure}{C\arabic{figure}}
\setcounter{figure}{0}
\subsection{Experiments behind Fig.~\ref{Sensitivity}}
\begin{figure}
 \begin{center}
  \includegraphics[width=1.00\columnwidth]{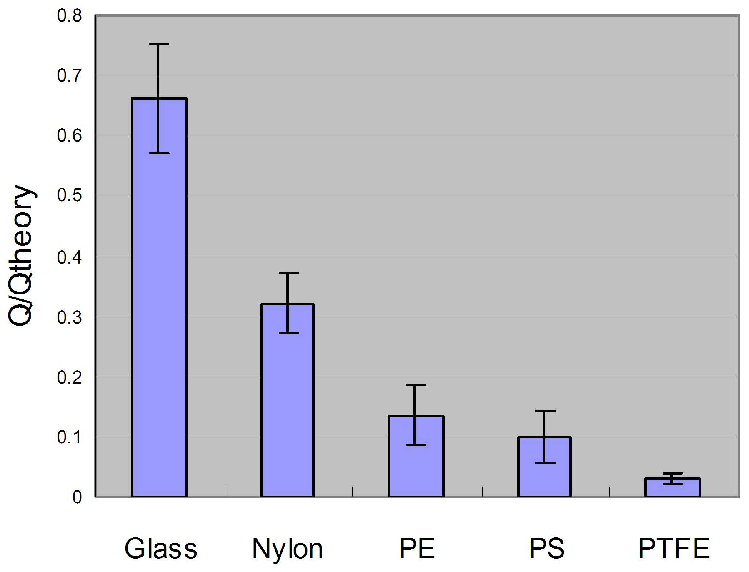}
 \end{center}
 \caption{Charge of spherical beads of different dielectric materials after contact with the electrode.}
 \label{charge}
\end{figure}

Fig.~\ref{Sensitivity} and its insets contains three types of information. First, it shows the contact angle between a $1\mu$l water drop and the insulator surface for different dielectric materials measured with an Optical Contact Angle Measuring Device. Second, it shows the average thickness of the water film for silicon oxide and PTFE as function of air humidity measured with an ellipsometer. Third, it shows the difficulty of charge transfer for different dielectric materials obtained from qualitative observations of the bouncing behavior of objects of different dielectric materials using the experimental setup of experiments~1, 2, and 3 described above (see Supplementary Movie 2). These qualitative observations were further backed up by quantitative measurements of the charge of spherical glass, Nylon, PE, PS, and PTFE beads after contact with the electrodes ($E=2.5$kV/cm). In fact, Fig.~\ref{charge} shows the portion of the charge
\begin{eqnarray}
 Q_\mathrm{theory}=\frac{\pi^2}{6}4\pi r^2\epsilon_o\epsilon_rE
\end{eqnarray}
an ideally conducting sphere with radius $r$ would obtain after contact with the electrodes, where $\epsilon_o$ is the permittivity of vacuum and $\epsilon_r$ the relative permittivity of silicone oil.
It can be seen that the more hydrophilic the material the larger is the charge after contact with the electrodes (cf. Fig.~\ref{Sensitivity}).

\subsection{Water adsorption from a humid atmosphere}
The amount of water adsorbed by an insulator from a humid atmosphere depends not only on the degree of humidity, but also on the chemical properties of the insulator surface. For instance, ionic insulators, such as NaCl and mica, adsorb water due to electrostatic interactions between the electric fields generated by the ions of the insulator and the dipole and higher moments of the water molecule \cite{Enkgvist_and_Stone_1999}, while hydrophilic covalent insulators, such as glass and quartz, adsorb water due to the generation of strong hydronium bonds \cite{Asay_and_Kim_2005}. Hydrophobic insulators, such as PE, PS, and PTFE, which do not adsorb water by either of these mechanisms, can still adsorb water due to ubiquitous dispersion interactions \cite{Asay_and_Kim_2005}, however to a much smaller extent than with the other two mechanisms.

\renewcommand{\thefigure}{E\arabic{figure}}
\setcounter{figure}{0}
\subsection{Possible mechanism for contact electrification of hydrophobic insulators}
\begin{figure}
 \begin{center}
  \includegraphics[width=1.00\columnwidth]{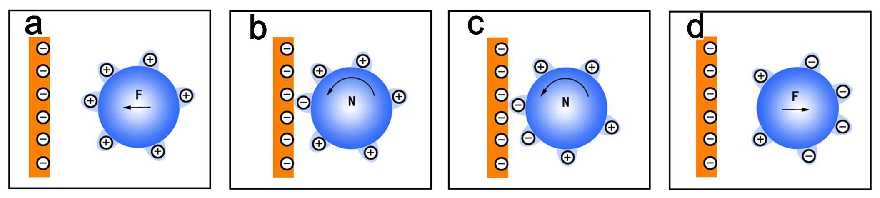}
 \end{center}
 \caption{(a-d) Charging mechanism of a hydrophobic particle in contact with the cathode.}
 \label{Chargemechanismhydrophob}
\end{figure}

Even though contact electrification was strongly resisted by the hydrophobic insulators made of PE and PS, it still occurred. Here we hypothesize how the charge transfer mechanism might have looked like in case of an insulator-electrode contact. First, we note that the size of the water islands on the surfaces of hydrophobic insulators is much smaller than on hydrophilic insulators, and much less water vapor is located near the insulator surface. This makes charge transport from water island to water island via the aforementioned exchange with surrounding water vapor much more difficult for hydrophobic particles than for hydrophilic particles. We illustrated this in Fig.~\ref{Chargemechanismhydrophob} through drawing isolated water islands, while the water film on hydrophilic particles was drawn as continuous in Figs.~\ref{Efield}d-f to illustrate that charge transport along it is easily possible. These water islands, however, still contain small amounts of excess charge \cite{Santos_et_al_2011} and are either positively or negatively charged (for illustration purposes, the charges are only positive in Fig.~\ref{Chargemechanismhydrophob}a). On contact with the cathode, the water island at the contact point becomes equipotential with it (cf. Fig.~\ref{Chargemechanismhydrophob}b). Moreover, asymmetries in the charges of the water islands or in the particle shape induce torsion, and the particle thus rolls under its action (cf. Fig.~\ref{Chargemechanismhydrophob}c). During this rolling process, more and more isolated adsorbed water domains become equipotential with the cathode, and the particles acquire thus more negative charges (cf. Fig.~\ref{Chargemechanismhydrophob}c). Once the net charge of the insulator becomes negative, the particle separates from the cathode (cf. Fig.~\ref{Chargemechanismhydrophob}d).

\section*{Acknowledgements}
We acknowledge support from grants NSFC No. 51275529, NSFC No. 41350110226, Science and Technology Development Plan of Qingdao City No. 12-1-4-7-(2)-jch, and Taishan Scholar Construction Project of Shandong Province No. TS20110823. We thank Fernando Galembeck for discussions.


\begin{thebibliography}{1}
%
\bibitem{Shinbrot_and_Herrmann_2008} Shinbrot, T. \& Herrmann, H. J., Granular matter: Static in motion. \textit{Nature} {\bf{451}}, 773-774 (2008).
%
\bibitem{Lacks_2010} Lacks, D. J., Particle clouds frictile attraction. \textit{Nat. Phys.} {\bf{6}}, 324-325 (2010).
%
\bibitem{Kok_and_Renno_2008} Kok, J. F. \& Renno, N. O., Electrostatics in wind-blown sand. \textit{Phys. Rev. Let.} {\bf{100}}, 014501 (2008).
%
\bibitem{Merrison_2012} Merrison, J. P., Sand transport, erosion and granular electrification. \textit{Aeolian. Res.} {\bf{4}}, 1 (2012).
%
\bibitem{McNutt_and_Davis_2000} McNutt, S. R. \& Davis, C. M., Lightning associated with the 1992 eruptions of crater peak, mount spurr volcano, Alaska. \textit{J. Volcanol. Geoth. Res.} {\bf{102}}, 45-65 (2000).
%
\bibitem{Abbasi_and_Abbasi_2007} Abbasi, T. \& Abbasi, S. A., Dust explosions-Cases, causes, consequences, and control. \textit{J. Hazard. Mater.} {\bf{140}}, 7-44 (2007).
%
\bibitem{Schein_2009} Schein, L., Comparison of toner adhesion theories. \textit{J. Imaging. Sci. Technol.} {\bf{53}}, 10506-1 (2009).
%
\bibitem{Polsen_et_al_2013} Polsen, E. S., Stevens, A. G. \& Hart, A. J., Laser printing of nanoparticle toner enables digital control of micropatterned carbon nanotube growth. \textit{ACS Appl. Mater. Interfaces} {\bf{5}}, 3656-3662 (2013).
%
\bibitem{Kawamoto_2007} Kawamoto, H., {\em{Electronic circuit printing, 3D printing and film formation utilizing electrostatic inkjet technology.}} In International Conference on Digital Printing Technologies vol. 23, p. 961 (2007).
%
\bibitem{Calvert_2001} Calvert, P., Inkjet printing for materials and devices. \textit{Chem. Mater.} {\bf{13}}, 3299-3305 (2001).
%
\bibitem{Kwetkus_1998} Kwetkus, B. A., Particle triboelectrification and its use in the electrostatic separation process. \textit{Particul. Sci. Technol.} {\bf{16}}, 55 (1998).
%
\bibitem{Castle_2008} Castle, G. S. P., {\em{Contact charging between particles; some current understanding.}} In Proceedings of the ESA Annual Meeting on Electrostatics 2008, p. M1 (2008).
%
\bibitem{McCarty_and_Whitesides_2008} McCarty, L. S. \& Whitesides, G. M., Electrostatic charging due to separation of ions at interfaces: contact electrification of ionic electrets. \textit{Angew. Chem. Int. Ed.} {\bf{47}}, 2188-2207 (2008).
%
\bibitem{Lacks_and_Sankaran_2011} Lacks, D. J. \& Sankaran, R. M., Contact electrification of insulating materials. \textit{J. Phys. D: Appl. Phys.} {\bf{44}}, 453001 (2011).
%
\bibitem{Pahtz_et_al_2010} T. P\"ahtz, H. J. Herrmann, T. Shinbrot, Why do particle clouds generate electric charges? \textit{Nat. Phys.} {\bf{6}}, 364-368 (2010).
%
\bibitem{Schein_2007} Schein, L. B., Recent progress and continuing puzzles in electrostatics. \textit{Science} {\bf{316}}, 1572-1573 (2007).
%
\bibitem{Liu_and_Bard_2008} Liu, C. \& Bard, A. J., Electrostatic electrochemistry at insulators. \textit{Nat. Mater.} {\bf{7}}, 505-509 (2008).
%
\bibitem{Diaz_and_Felix-Navarro_2004} Diaz, A. F. \& Felix-Navarro, R. M., A semi-quantitative tribo-electric series for polymeric materials: the influence of chemical structure and properties. \textit{J. Electrostat.} {\bf{62}}, 277-290 (2004).
%
\bibitem{Baytekin_et_al_2011a} Baytekin, H. T. {\em{et al.}}, The mosaic of surface charge in contact electrification. \textit{Science} {\bf{333}}, 308-312 (2011).
%
\bibitem{Gu_et_al_2013} Gu, Z., Wei, W., Su, J. \& Yu, C. W., The role of water content in triboelectric charging of wind-blown sand. \textit{Sci. Rep.} {\bf{3}}, 1337 (2013).
%
\bibitem{Ducati_et_al_2010} Ducati, T. R., Sim\~{o}es, L. H. \& Galembeck, F., Charge partitioning at gas-solid interfaces: Humidity causes electricity buildup on metals. \textit{Langmuir} {\bf{26}}, 13763-13766 (2010).
%
\bibitem{Williams_2012a} Williams, M. W., What creates static electricity? \textit{Am. Sci.} {\bf{100}}, 316-323 (2012).
%
\bibitem{Burgo_et_al_2013} Burgo, T. A. L. {\em{et al.}}, Friction coefficient dependence on electrostatic tribocharging. \textit{Sci. Rep.} {\bf{3}}, 2384 (2013).
%
\bibitem{Williams_2012b} Williams, M. W., Triboelectric charging of insulators-mass transfer versus electrons/ions. \textit{J. Electrostat.} {\bf{70}}, 233-234 (2012).
%
\bibitem{Knorr_2011} Knorr, N., Squeezing out hydrated protons: low-frictional-energy triboelectric insulator charging on a microscopic scale. \textit{AIP Adv.} {\bf{1}}, 022119-022119-22 (2011).
%
\bibitem{Kamra_1972} Kamra, A. K., Visual observation of electric sparks on gypsum dunes. \textit{Nature} {\bf{240}}, 143-144 (1972).
%
\bibitem{Xu_et_al_2010} Xu, K., Cao, P. \& Health, J. R., Graphene Visualizes the First Water Adlayers on Mica at Ambient Conditions. \textit{Science} {\bf{329}}, 1188-1191 (2010).
%
\bibitem{Carrasco_et_al_2012} Carrasco, J., Hodgson, A., Michaelides, A., A molecular perspective of water at metal interfaces. \textit{Nat. Mater.} {\bf{11}}, 667 (2012).
%
\bibitem{Burgo_et_al_2011} Burgo T. A. L., Rezende C. A., Bertazzo S., Galembeck A., Galembeck F., Electric potential decay on polyethylene: role of atmospheric water on electric charge build–up and dissipation. \textit{J. Electrostat.} {\bf{69}}, 401 (2011).
%
\bibitem{Wiles_et_al_2004} Wiles J. A., Fialkowski M., Radowski M. R., Whitesides G. M., Grzybowski B. A., Effects of Surface Modification and Moisture on the Rates of Charge Transfer between Metals and Organic Materials. J. Phys. Chem. B {\bf{108}}, 20296 (2004).
%
\bibitem{Gouveia_and_Galembeck_2009} Gouveia R. F., Galembeck F., Electrostatic charging of hydrophilic particles due to water adsorption. J. Am. Chem. Soc. {\bf{131}}, 11381 (2009).
%
\bibitem{Pence_et _al_1994} Pence, S., Novotny, V. J., Diaz, A. F., Effect of Surface Moisture on Contact Charge of Polymers Containing Ions., Langmuir {\bf{10}}, 592 (1994).
%
\bibitem{Baytekin_et_al_2011b} Baytekin, H. T., Baytekin, B., Soh, S., Grzybowski, B. A., Is Water Necessary for Contact Electrification?, Angew. Chem. {\bf{50}}, 6766 (2011).
%
\bibitem{Zheng_et_al_2014} Zheng X., Zhang R., Huang H., Theoretical modeling of relative humidity on contact electrification of sand particles., Sci. Rep. {\bf{4}}, 4399 (2014).
%
\bibitem{Kok_and_Renno_2006} Kok, J. F. \& Renno, N. O., Enhancement of the emission of mineral dust aerosols by electric forces. \textit{Geophys. Res. Lett.} {\bf{33}}, L19S10 (2006).
%
\bibitem{Sow_et_al_2013} Sow, M., Akande, A. R., Robinson, K. S., Sankaran, R. M. \& Lacks, D. J., The Role of Humidity on the Lift-off of Particles in Electric Fields. \textit{J. Braz. Chem. Soc.} {\bf{24}}, 273 (2013).
%
\bibitem{Lebedev_Skalskays_1962} Levedev, N. N. \& Skalskaya, I. P., Force acting on a conducting sphere in the field of a parallel plate condenser. \textit{Sov. Phys. Tech. Phys., Engl. Transl.} {\bf{7}}, 268 (1962).
%
\bibitem{Ristenpart_et_al_2009} Ristenpart, W. D. {\em{et al.}}, Non-coalescence of oppositely charged drops. \textit{Nature} {\bf{461}}, 377-380 (2009).
%
\bibitem{Chakrapani_et_al_2007} Chakrapani, V. {\em{et al.}}, Charge transfer equilibria between diamond and an aqueous oxygen electrochemical redox couple. \textit{Science} {\bf{318}}, 1424-1430 (2007).
%
\bibitem{Ewing_2006} Ewing, G. E., Ambient thin film water on insulator surfaces. \textit{Chem. Rev.} {\bf{106}}, 1511-1526 (2006).
%
\bibitem{Siu_et_al_2014} Siu, T., Cotton, J., Mattson, G. \& Shinbrot, T., Self-sustaining charging of identical colliding particles. \textit{Phys. Rev. E} {\bf{89}}, 052208 (2014).
%
\bibitem{Rezende_et_al_2009} Rezende C. A., Gouveia R. F., da Silva M. A., Galembeck F., Detection of charge distributions in
insulator surfaces. \textit{Journal of Physics: Condensed Matter} {\bf{21}}, 263002 (2009).
%
\bibitem{Im_et_al_2012} Im D. J., Ahn M. M., Yoo B. S., et al., Discrete electrostatic charge transfer by the electrophoresis of a charged droplet in a dielectric liquid. \textit{Langmuir} {\bf{28}}, 11656 (2012).
%
\bibitem{Im_et_al_2011} Im D. J., Noh J., Moon D., et al., Electrophoresis of a charged droplet in a dielectric liquid for droplet actuation. \textit{Analytical Chemistry} {\bf{83}}, 5168 (2011).
%
\bibitem{Enkgvist_and_Stone_1999} Enkgvist, O. \& Stone, A. J.,  Intermolecular potentials and low temperature structures. \textit{J. Chem. Phys.} {\bf{110}}, 12089 (1999).
%
\bibitem{Asay_and_Kim_2005} Asay, D. B. \& Kim, S. H., Evolution of the adsorbed water layer structure on silicon oxide at room temperature. \textit{J. Phys. Chem.} B {\bf{109}}, 16760 (2005).
%
\bibitem{Santos_et_al_2011} Santos, L. P., Ducati, R. T. R. D., Balestrin, L. B. S., Galembeck, F., Water with excess electric charge. \textit{J. Phys. Chem.} C {\bf{115}}, 11226 (2011).
%
\end{thebibliography}
\end{document}